\documentclass[a4paper,11pt]{article}
\usepackage{jinstpub} % for details on the use of the package, please see the JINST-author-manual
\usepackage{lineno}
\usepackage{subcaption}

\usepackage{makecell}

\title{\boldmath PMT calibration for the JSNS$^2$-II far detector with an embedded LED system }

\author[a]{Jisu Park,} % OK
\author[b]{M. K. Cheoun,} % OK
\author[c]{J. H. Choi,} % OK
\author[a]{J. Y. Choi,} % OK
\author[d,g]{T. Dodo,} % OK
\author[e]{J. Goh,} % OK
\author[f]{M. Harada,} % OK
\author[g,f]{S. Hasegawa,} % OK
\author[e]{W. Hwang,} % OK
\author[h]{T. Iida,} % OK
\author[i]{H. I. Jang,} % OK
\author[j]{J. S. Jang,} % OK
\author[a]{K. K. Joo,} % OK
\author[a]{D. E. Jung,} % OK
\author[k]{S. K. Kang,} % OK
\author[f]{Y. Kasugai,} %OK
\author[l]{T. Kawasaki,} % OK
\author[a]{E. M. Kim,} % OK
\author[m]{S. B. Kim,} % OK
\author[a]{S. Y. Kim,} % OK
\author[f]{H. Kinoshita,} % OK
\author[l]{T. Konno,} % OK
\author[n]{D. H. Lee,} % OK
\author[o]{C. Little,} % OK
\author[n]{T. Maruyama,} % OK
\author[o]{E. Marzec,} % OK
\author[f]{S. Masuda,} % OK
\author[f]{S. Meigo,} % OK
\author[a]{D. H. Moon,} % OK
\author[p]{T. Nakano,} % OK
\author[q]{M. Niiyama,} % OK
\author[n]{K. Nishikawa,}  % OK
\author[c]{M. Y. Pac,} % OK
\author[r]{B. J. Park,} % OK
\author[a]{H. W. Park,} % OK
\author[b]{J. B. Park,} % OK
\author[r]{J. S. Park,} % OK
\author[a]{R. G. Park,} % OK
\author[s]{S. J. M. Peeters,} % OK
\author[t,u]{C. Rott,} % OK
\author[r]{J. W. Ryu,} % OK
\author[f]{K. Sakai,} % OK
\author[f]{S. Sakamoto,} % OK
\author[p]{T. Shima,} % OK
\author[n]{C. D. Shin, \footnote{Corresponding author.}} % OK
\author[o]{J. Spitz,} % OK
\author[d]{F. Suekane,} % OK
\author[p]{Y. Sugaya,} % OK
\author[f]{K. Suzuya,} % OK
\author[h]{Y. Takeuchi,} % OK
\author[m]{W. Wang,} % OK
\author[m]{W. Wei,} % OK
\author[f]{Y. Yamaguchi,} % OK
\author[v]{M. Yeh,} % OK
\author[c]{I. S. Yeo,} % OK
\author[u]{and I. Yu} % OK

\affiliation[a]{Department of Physics, Chonnam National University,\\ 77, Yongbong-ro, Buk-gu, Gwangju, 61186, Korea}
\affiliation[b]{Department of Physics, Soongsil University,\\ 369 Sangdo-ro, Dongjak-gu, Seoul, 06978, Korea}
\affiliation[c]{Laboratory for High Energy Physics, Dongshin University,\\ 67, Dongshindae-gil, Naju-si, Jeollanam-do, 58245, Korea}
\affiliation[d]{Research Center for Neutrino Science, Tohoku University,\\ 6-3 Azaaoba, Aramaki, Aoba-ku, Sendai 980-8578, Japan}
\affiliation[e]{Department of Physics, Kyung Hee University,\\ 26, Kyungheedae-ro, Dongdaemun-gu, Seoul 02447, Korea}
\affiliation[f]{J-PARC Center, JAEA,\\ 2-4 Shirakata, Tokai-mura, Naka-gun, Ibaraki 319-1195, Japan}
\affiliation[g]{Advanced Science Research Center, JAEA,\\ 2-4 Shirakata, Tokai-mura, Naka-gun, Ibaraki 319-1195, Japan}
\affiliation[h]{Faculty of Pure and Applied Sciences, University of Tsukuba,\\ Tennodai 1-1-1, Tsukuba, Ibaraki, 305-8571, Japan}
\affiliation[i]{Department of Fire Safety, Seoyeong University,\\ 1 Seogang-ro, Buk-gu, Gwangju, 61268, Korea}
\affiliation[j]{GIST College, Gwangju Institute of Science and Technology,\\ 123 Cheomdangwagi-ro, Buk-gu, Gwangju, 61005, Korea}
\affiliation[k]{School of Liberal Arts, Seoul National University of Science and Technology,\\ 232 Gongneung-ro, Nowon-gu, Seoul, 139-743, Korea}
\affiliation[l]{Department of Physics, Kitasato University,\\ 1 Chome-15-1 Kitazato, Minami Ward, Sagamihara, Kanagawa, 252-0329, Japan}
\affiliation[m]{School of Physics, Sun Yat-sen (Zhongshan) University,\\ Haizhu District, Guangzhou, 510275, China}
\affiliation[n]{High Energy Accelerator Research Organization (KEK),\\ 1-1 Oho, Tsukuba, Ibaraki, 305-0801, Japan}
\affiliation[o]{University of Michigan,\\ 500 S. State Street, Ann Arbor, MI 48109, U.S.A.}
\affiliation[p]{Research Center for Nuclear Physics, Osaka University,\\ 10-1 Mihogaoka, Ibaraki, Osaka, 567-0047, Japan}
\affiliation[q]{Department of Physics, Kyoto Sangyo University,\\ Motoyama, Kamigamo, Kita-Ku, Kyoto-City, 603-8555, Japan}
\affiliation[r]{Department of Physics, Kyungpook National University,\\ 80 Daehak-ro, Buk-gu, Daegu, 41566, Korea}
\affiliation[s]{Department of Physics and Astronomy, University of Sussex,\\ Falmer, Brighton, BN1 9RH, U.K.}
\affiliation[t]{Department of Physics and Astronomy, University of Utah,\\ 201 Presidents' Cir, Salt Lake City, UT 84112, U.S.A}
\affiliation[u]{Department of Physics, Sungkyunkwan University,\\ 2066, Seobu-ro, Jangan-gu, Suwon-si, Gyeonggi-do, 16419, Korea}
\affiliation[v]{Brookhaven National Laboratory,\\ Upton, NY 11973-5000, U.S.A.}

\emailAdd{cdshin@post.kek.jp}

\abstract
{
	The JSNS$^{2}$-II (the second phase of JSNS$^2$, J-PARC Sterile Neutrino Search at J-PARC Spallation Neutron Source) is an 
    experiment aimed at searching for sterile neutrinos. This experiment has entered its second phase,
    employing two liquid scintillator detectors located at near and far positions from the neutrino source.
    Recently, the far detector of the experiment has been completed and is currently in the 
    calibration phase. This paper presents a detailed description of the calibration 
    process utilizing the LED system. The LED system of the far detector uses two Ultra-Violet (UV) LEDs, which are effective in calibrating all of PMTs at once. The UV light is converted into the visible
    light wavelengths inside liquid scintillator via the wavelength shifters, providing
    pseudo-isotropic light. The properties of all functioning Photo-Multiplier-Tubes (PMTs) to 
    detect the neutrino events in the far detector, such as gain, its dependence of supplied High Voltage (HV), and Peak-to-Valley (PV) were calibrated. To achieve a good energy resolution for physics events, up to 10\% of the relative gain adjustment is required for all functioning~PMTs. This will be achieved using the measured HV curves and the LED calibration. The Peak-to-Valley (PV) ratio values are the similar to those from the production company, which distinguish the single photo-electron signal from the pedestal. Additionally, the precision of PMT signal timing is measured to be 2.1~ns, meeting the event reconstruction requirement of 10~ns.
}

\keywords{Neutrino detectors; Scintillators; scintillation and light emission processes (solid, gas
and liquid scintillators); Gamma detectors (scintillators, CZT, HPGe, HgI etc)}

\begin{document}
	\maketitle
	\flushbottom
	
	\section{Introduction}
	\label{sec:intro}

    The JSNS$^2$-II is the second phase of the
    JSNS$^{2}$ (J-PARC Sterile Neutrino Search at J-PARC Spallation Neutron Source) experiment.
    The experiment aims to search for neutrino oscillations ($\bar{\nu}_{\mu} \to \bar{\nu}_{e}$) using short baselines (24 and 48 meters from the neutrino source) and two liquid scintillator detectors. Neutrino oscillations with short baselines have suggested the existence of the sterile neutrinos, as indicated by several experiments~\cite{cite:LSND,CITE:MiniBooNE,CITE:BEST,CITE:Reactor}.  
    The intense $\bar{\nu}_{\mu}$ flux is produced via $\mu^{+}$ decay-at-rest at the J-PARC Material and Life Science Experimental Facility (MLF) mercury target when the 1 MW beam of 3~GeV protons impacts the target. These 3~GeV protons are accelerated by the J-PARC Rapid Cycle Synchrotron (RCS). The delivered protons come in two~bunches at a 25~Hz repetition rate. Each bunch has a width of 100~ns, with the two bunches seperated by 600~ns.
    JSNS$^2$~\cite{JSNS2_proposal} was proposed in 2013,
    and its second phase, JSNS$^2$-II~\cite{JSNS2_II_Proposal}, was proposed in 2021.
    
\begin{figure}[htbp]
		\centering
		\includegraphics[width=.58\textwidth]{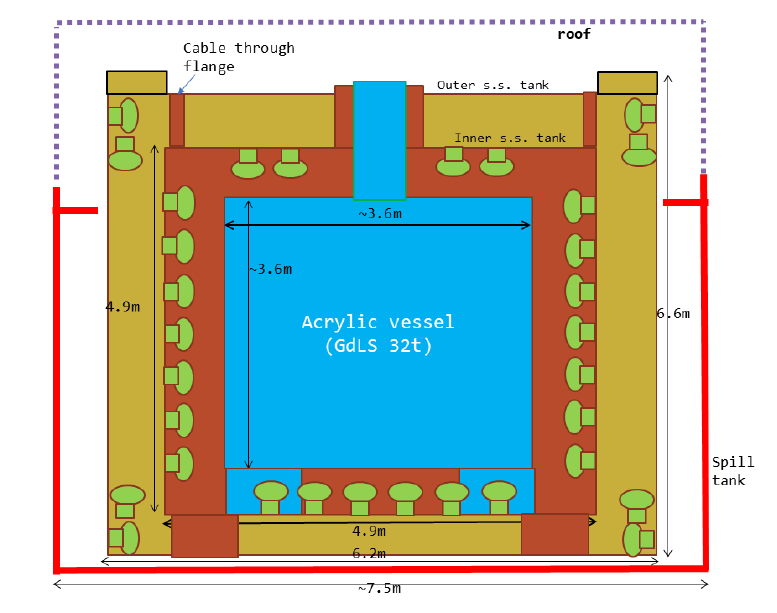}
		\caption{Structure of the JSNS$^{2}$-II detector}
		\label{fig:Detector}
	\end{figure}

    The near detector, located 24 meters from the neutrino source on the third floor of the J-PARC MLF, began data taking in 2020. The far detector, 
    situated 48 meters away outside the MLF building, was recently completed. 
    The conceptual design of the far detector is illustrated in Figure~\ref{fig:Detector}. 
    
    The far detector consists of three layers, from the innermost to the outermost: a target, a gamma-catcher, and a veto region. The target region contains an acrylic 
    vessel~\cite{CITE:JSNS2_II_acrylic} with 32~tonnes of Gadolinium-loaded liquid scintillator (Gd-LS) for neutrino detection, while the other two layers house 131~tonnes of Gadolinium-unloaded liquid scintillator (pureLS). To detect neutrino events, 172 Photo-Multiplier-Tubes (PMTs) surround the acrylic vessel within the inner stainless steel tank, while 48 PMTs are positioned 
    in the veto region, inside the outer stainless steel tank, to capture incoming particles from outside of the detector. 
    
    The PMT signals are digitized and recorded using CAEN V1730SB modules, which offer 
    a 14~bits resolution and 500~MHz sampling rate~\cite{CITE:CAEN}. The full dynamic range of the boards is 2~$V$.
    
    This paper describes the calibration process for the all functioning inner PMTs of the far detector utilizing an LED system.     
    
	\section{Requirements for the detector and calibration}
	\label{sec:requirement}

To search for the sterile neutrinos, the experiment looks for neutrino oscillation 
from $\bar{\nu}_{\mu}$ to $\bar{\nu}_{e}$ with short baselines of 24 and 48 meters. The beam intrinsic $\bar{\nu}_{e}$ background is also produced via the $\pi^{-} \to \mu^{-}$ chain
in the mercury target although the fraction of this chain is suppressed by a factor of 1000 compared to the number of produced 
particles at the target due to the $\pi^{-}$ and $\mu^{-}$ being captured on the mercury nuclei. However, the energy spectra difference between the oscillated signal and the intrinsic background is used for the sterile neutrino search~\cite{JSNS2_proposal} due to the small oscillation mixing
angles for the searches.

To distinguish the oscillated signals from the background, both energy calibration and energy resolution of the detectors are crucial. The energy and vertex will be reconstructed by a maximum likelihood technique using PMT charge information ~\cite{CITE:Johnathon, CITE:Cf}.
To achieve good differentiation, an energy resolution of less than 3\% at 53 MeV, 
which is the endpoint of the signal, is necessary.

The detector placed at 24~m met the 
requirement~\cite{CITE:Cf, CITE:PRL}, and the gain matching of the PMTs is required to be less than 10\% with the supplied high voltages. 
To meet this requirement, precise gain curves as a function of high voltages for each PMT are necessary. 

The pedestal fluctuation and typical single photo-electron (s.p.e.) signal must be well-separated to identify the low-energy signals accurately. To achieve good pulse detection, the peak-to-valley ratios of charge distributions around zero and s.p.e. must to be more than 
1.5, which is the minimum specification of the Hamamatsu photonics~\cite{CITE:Hamamatsu}. 

Offline event building requires that the timing of each PMT hit is concentrated within a 60~ns time window. To perform event building correctly with all PMTs, the timing calibration of 
each PMT must be within 10~ns. As shown in
Figure~\ref{fig:Detector}, the largest distance inside the gamma-catcher region is about 7~m; therefore, the 35~ns difference could
be observered in the PMT timing data. 

\section{PMTs}

The JSNS$^2$-II far detector uses R7081 10-inch PMTs made by Hamamatsu Photonics. All 172~PMTs in the gamma-catcher region were contributed by the Double-Chooz experiment. Figure~\ref{Fig:ChargeMap} shows the locations of the inner PMTs.  

Thirty veto PMTs were newly produced, while 18 PMTs were reused from the Double-Chooz experiment. The operating High Voltage (HV) points in JSNS$^2$-II differ from those in the Double-Chooz experiment; therefore, this calibration is crucial.

The general properties of the R7081 10-inch PMT, as detailed in references~\cite{CITE:Hamamatsu, CITE:Matsubara} are summarized in Table~\ref{Tab:PMT}.    

 \begin{table}[h]
   \begin{center}
      \begin{tabular*}{0.81\textwidth}{ c|c|c|c|c|c} 
        \hline
        \makecell{Applied voltage \\ for 10$^7$ gain (V)} &
        \multicolumn{2}{c|}{PV ratio} & 
        Transit time spread (ns) & 
        \multicolumn{2}{c}{\makecell{Dark count (s$^{-1}$)\\ (After 15 hours)}} \\ \hline
        Typ. & Min. & Typ. & Typ. & \hspace{1mm} Typ.\hspace{1mm} & Max.\\ \hline
        1500 & 1.5 & 2.8 & 3.4 & 7000 & 15000 \\ \hline
      \end{tabular*}
      \caption{Typical properties of R7081 10-inch PMT~\cite{CITE:Hamamatsu}. }
      \label{Tab:PMT}
   \end{center}
\end{table}

In this manuscript, we report the HV curves, Peak-to-Valley (PV) ratios and timing determination capability obtained from the in-situ calibration of the far detector using an LED system.

\section{LED system}

\subsection{System overview}

\begin{figure}[!bt]
	\centering
	\includegraphics[width=.6\textwidth]{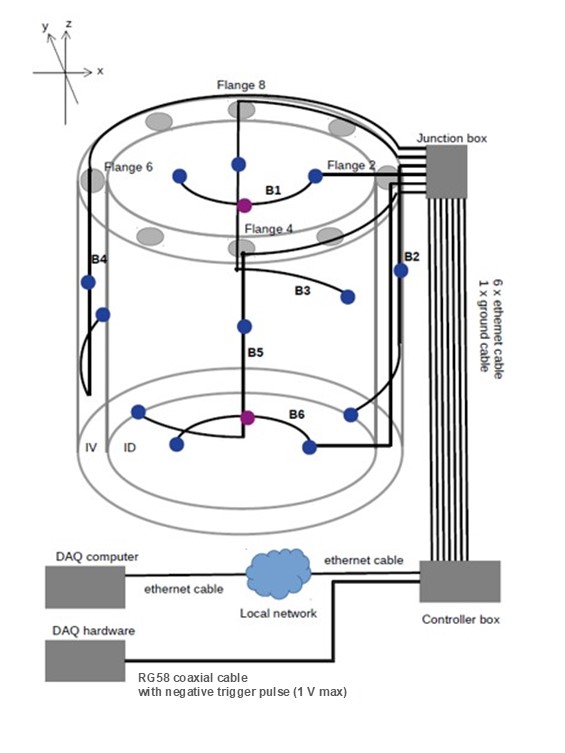}
	\caption{An overview of the JSNS$^{2}$-II LED calibration system. The LED drivers are connected via six branches, entering the detector through top flanges and located between the Inner Veto (IV) volume, the PMTs and the Inner Detector (ID) volume. The blue points indicate the location of the LED pulsers with 420~nm LEDs. The purple points indicated the location of the LED pulsers with 355~nm LEDs. The controller box is connected to the central data-acquisition (DAQ) computer via the local Ethernet network. The controller box also sends a trigger pulse in sync with the LED pulse to the DAQ hardware.}
	\label{fig:led_overview}
\end{figure}

The JSNS$^2$-II LED system (see Figure~\ref{fig:led_overview}) is based on an existing system developed by the University of Sussex and successfully used by SNO+\,\cite{bib:1} and Double Chooz\,\cite{bib:2}.
For JSNS$^2$-II, the system has been significantly improved, also by the University of Sussex. The main difference between the JSNS$^2$-II LED system and previous LED systems is the use of direct LED light inside the detector without the need to use a multimode polymethyl methacrylate (PMMA) optical fiber to guide the light into the detector. Both the near and far detectors of JSNS$^2$-II use the same LED calibration system.

\begin{figure}[!tb]
	\centering
	\includegraphics[width=.6\textwidth]{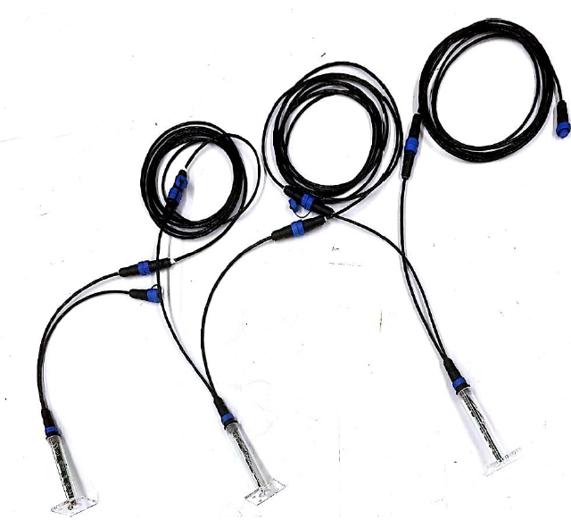}
	\caption{An image of one of the branches with three LED pulser daisy-chained on a single cable. All materials used are chemically compatible with LAB and the LED drivers are leak-tested and do not cause any electrical interference with the PMTs.}
	\label{fig:led_overview}
\end{figure}

\begin{figure}[!bt]
	\centering
	\includegraphics[width=.6\textwidth]{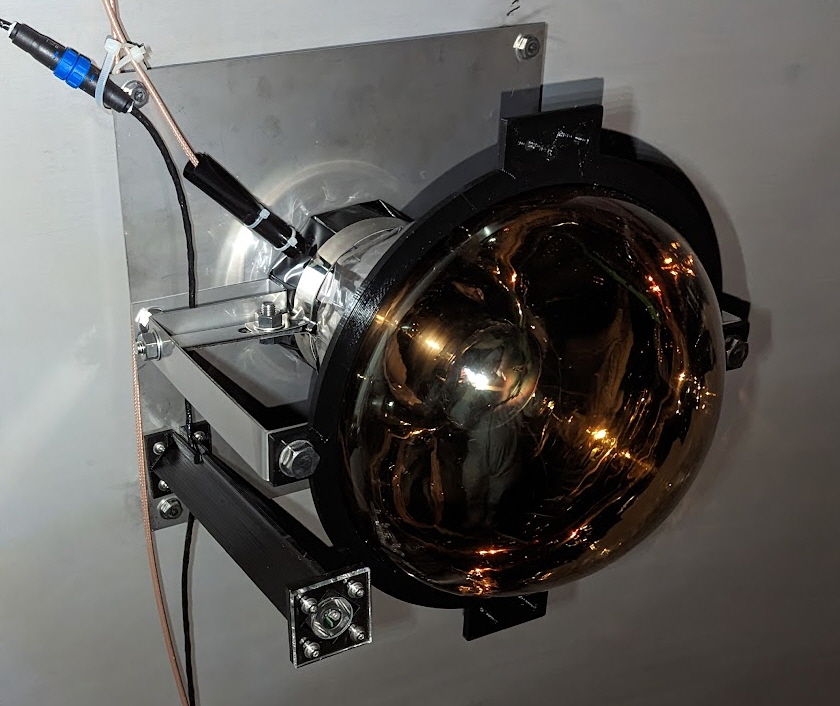}
	\caption{An image of LED on the PMT jig.}
	\label{fig:LED}
\end{figure}

Moving the LEDs to the inside of the detector has five important advantages. First, the calibration accuracy will be significantly improved compared to the traditional technique as the sub-nanosecond risetime of the pulse will not be degreaded by transmission through multimode fibers. Second, this gives us the capability to provide higher light intensity. Third, there is no bias on the timing properties as function of LED emission angle as introduced by the multimode PMMA fibers. Fourth, the device is suitable for the JSNS$^{2}$-II  detectors, because PMMA optical fibers are not compatible with Linear Alkyl Benzene (LAB). The JSNS$^2$-II LED system, on the other hand, can embed an LED with small driver board in the detector (see Figure~\ref{fig:led_overview}). The LED and drive board are encapsulated in a material compatible container made of acrylic. The encapsulated LEDs are mounted on the PMT jig using a 3D-printed LED holder as shown in the Figure~\ref{fig:LED}. Finally, directly submerging the pulsers allows the use of ultraviolet (UV) LEDs, as the wavelength is not transmitted by any significant length of PMMA fiber.

\begin{figure}[htbp]
  \centering
  \includegraphics[width=.85\textwidth]{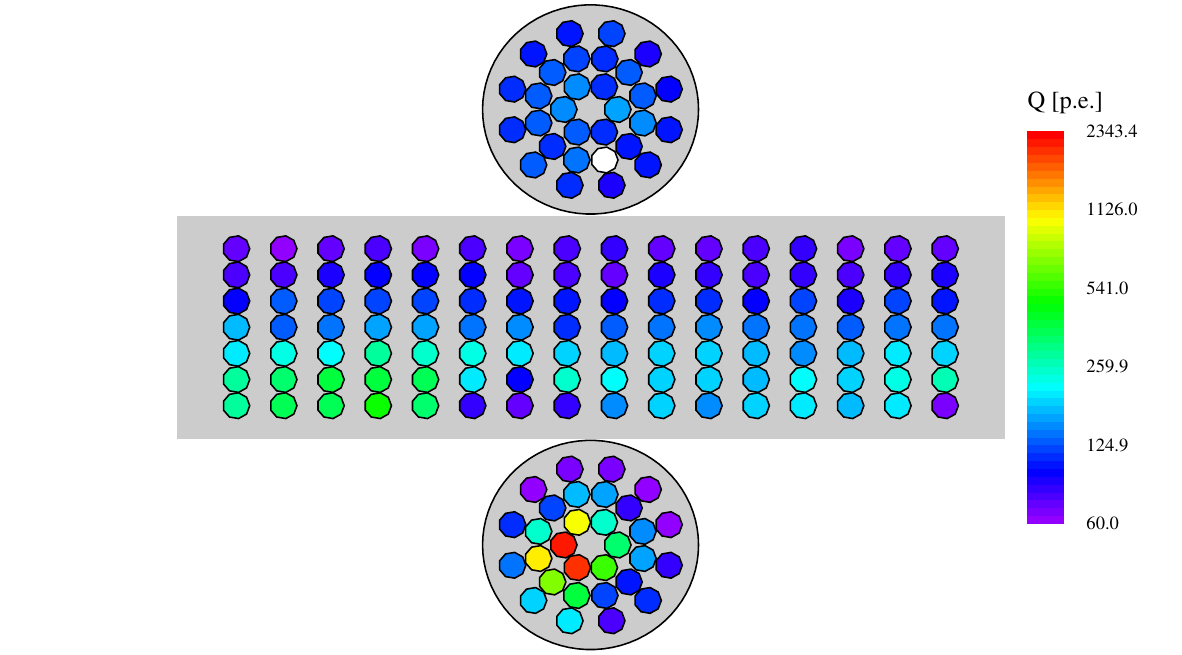}
  \caption{The relative charge map of the 172 PMTs using the UV LED located at the bottom of the detector. The white circle shows one dead PMT.}
   \label{Fig:ChargeMap}
\end{figure}

The 420~nm light does traverses the detector volume and is used at 12 points, ensuring full coverage of the detector. An UV LED with a wavelength 355~nm, used at 2 points (top and bottom) plays a crucial role for the JSNS$^2$ detectors, since the light from the LED is converted into observable (by the PMTs) light by wavelength shifting properties of the liquid scintillator. This indicates that the converted light is pseudo-isotropic, exhibiting minimal directional dependence in intensity, unlike standard LEDs. The 420 nm and 355 nm LEDs were used for gain and signal timing measurements, respectively. Figure~\ref{Fig:ChargeMap} illustrates the relative charge maps with the UV LED positioned at the bottom.
The UV LED is fixed near the PMT that exhibits the highest charge (indicated by the red circle in the figure). The LED emits pseudo-isotropic light. 
The specifications of the JSNS$^2$-II LED system are summarized in Table~\ref{fig:LED_specifications}. 

 %*** Table 1
 \begin{table}[htbp]
   \centering
   \begin{tabular}{ll} \hline
     Item &  Performances \\ \hline
     Wavelength & 355 and 420\,nm \\
     Timing profile of light pulse & 355\,nm: 0.40\,nsec (rise), 0.75\,nsec (fall), 0.79\,nsec  (width) \\
     & 420\,nm: 0.40\,nsec (rise), 0.63\,nsec (fall), 0.65\,nsec (width) \\
     Opening angle of LED light & 355\,nm: 30.2 $\pm$ 3.5\,degree\\
     & 420\,nm: 26.9 $\pm$ 1.8\,degree\\
     Light intensity &  Capability to provide (approximately) 100 to 1,000,000  photoelectrons \\
     Flashing rate & Up to 100\,kHz \\ 
     Trigger & Capability to produce/accept a TTL trigger \\ \hline
   \end{tabular}
   \caption{Major specification of the JSNS$^2$-II LED system.}
   \label{fig:LED_specifications}
 \end{table}

\subsection{System design}

The JSNS$^2$-II system control was modified to ensure synchronization of the timing between individual pulses. The system consists of two distinct components: the driver board, which contains the pulsing electronic circuit and LED, and the controller system, which interfaces with all driver boards and serves as the control interface.

The driver board's pulsing circuit is a Kapustinsky pulse circuit~\cite{bib:kap} featuring minor modifications developed for prototype LED driver circuits for the SNO+ project~\cite{bib:sussex}. 
The pulse intensity and trigger pulse signal are generated via an onboard microcontroller. 
Each driver board's microcontroller is programmed with a unique software image with a corresponding unique ID to allow them to be uniquely identified on a common communication bus. 
When generating a trigger pulse, one copy of the pulse signal is sent to an output buffer and back to the controller system for synchronization with the wider readout and calibration system. 
A secondary copy of the trigger pulse is routed through dual digital delay line Integrated Circuits (ICs).
These provide a coarse and fine level of delay tuning to the system to account for various cable lengths and other delays within system, allowing for output LED light to synchronize within the system. 
After this, the signal enters the pulse circuitry and generates a light pulse with the connected LED.

The controller system contains a custom control printed circuit board (PCB) with a Peripheral Interface Controller (PIC). 
The PIC provides a serial interface to allow for configuration messages to be transmitted to the various driver boards.
The synchronization pulse from each driver board when in operation is sent to the controller box where it pulses through a trigger pulse shaping circuit before being sent out to a Bayonet Neil-Concelman (BNC) connection on the controller case.
This allows for external systems such as Analogue-to-Digital converters (ADCs) to synchronize with the light pulse generated by the system. 

Finally, the controller box hosts a Raspberry Pi single-board computer. 
This hosts the JSNS$^2$-II LED system software, which contains various scripts for configuring pulser system via a standard Linux terminal. 
When using these tools, the configuration commands required are sent over a serial interface to the PIC and any response messages are sent back over the same interface. 
The controller box is connected to via an ethernet connection and therefore can be deployed via an Ethernet network for remote access.
    
\section{In-situ PMT calibrations using the LED system}
\subsection{PMT gain calibration}

\begin{figure}[htbp]
    \centering
    \includegraphics[width=.7\textwidth]{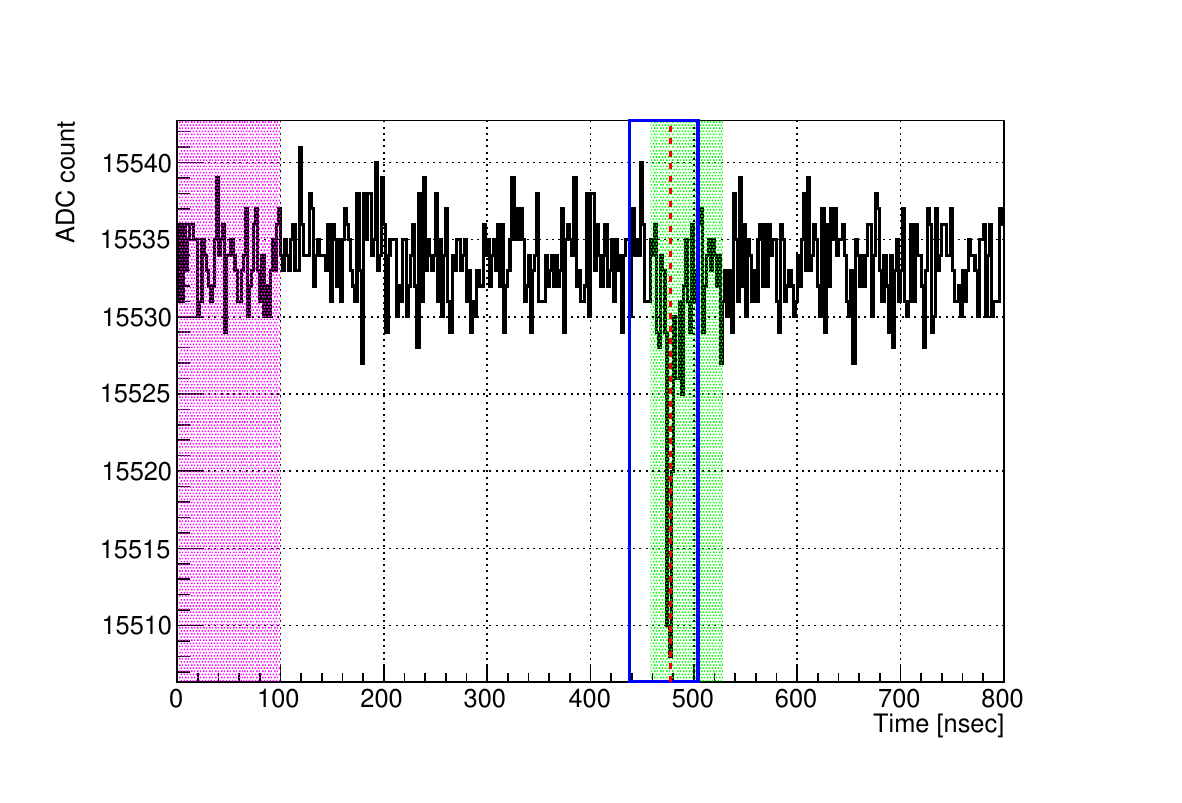}
    \caption{A typical PMT waveform from the LED calibration measured by the CAEN modules. The magenta box represents the pedestal calculation region, corresponding to the first 100~ns of the gate of the CAEN modules. The blue lines indicates the LED signal search range, determined by data with a high LED intensity. Once the pulse peak (red dashed line) is identified within the blue lines region, the green box, spanning from -20~ns to +50~ns relative to the peak, is defined as the charge integration range.}
    \label{Fig:WF}
\end{figure}

The advantage of using the UV LED is its ability to provide a good intensity for the single photoelectrons measurement for many PMTs at the same time due to the pseudo-isotropic light. For this calibration, the LED system sends trigger pulses to the CAEN modules at a rate of 200~Hz, while the PMT output is recorded within a 2~$\mu$s time window.
Figure~\ref{Fig:WF} shows the typical PMT waveform measured during the LED calibration. 
    
\begin{figure}[htbp]
    \centering
    \includegraphics[width=.7\textwidth]{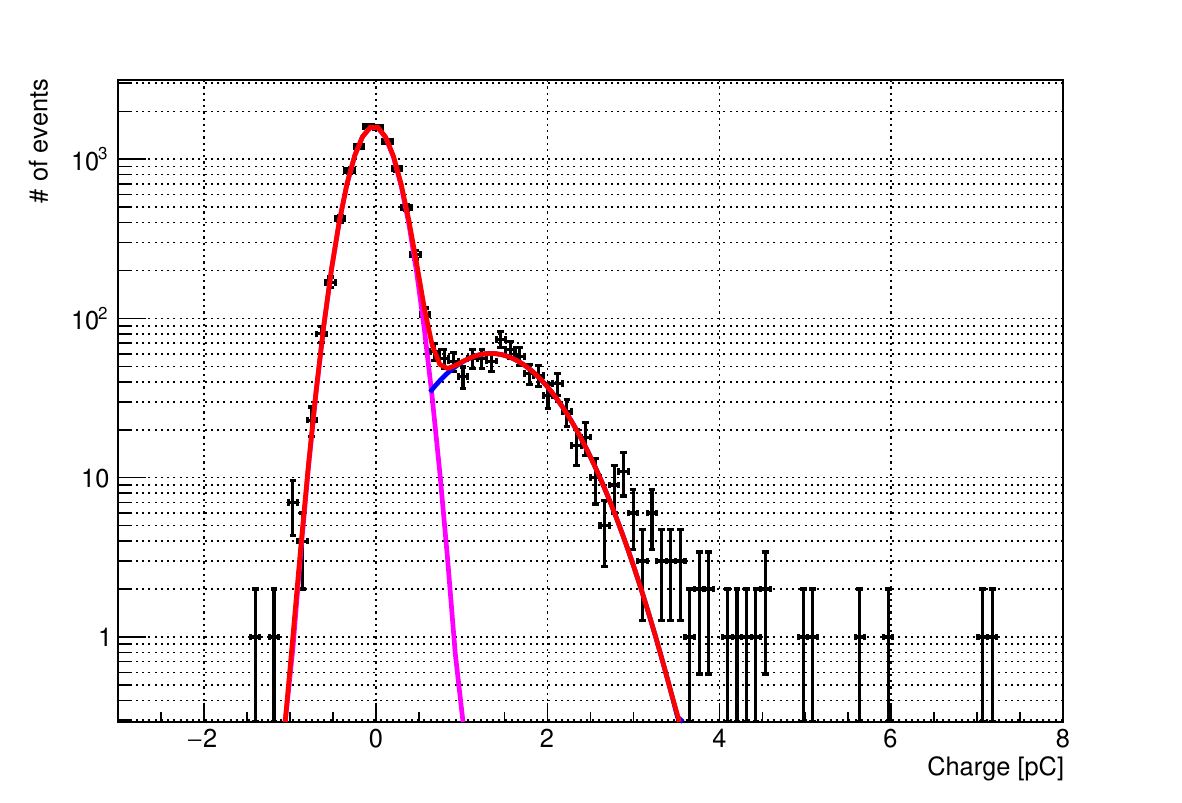}
    \caption{An example of the charge distribution measured by PMT during the LED run. }
    \label{Fig:SPE}
\end{figure}

The first 100~ns is used for the pedestal calculation while the CAEN modules' counts are accumulated in the specific time window of the LED timing trigger, then the CAEN modules' counts are converted into the pC.  
Figure~\ref{Fig:SPE} shows the charge distribution from one PMT using the UV LED. The clear single photoelectron (s.p.e.) is found. To calculate the peak position of the single photoelectron, a double-Gaussian fit function is adopted to model both the pedestal and the single photoelectron.
As shown in Figure~\ref{Fig:ChargeMap}, it is difficult to calibrate the PMTs at the bottom of the detector, therefore another LED was used to calibrate them.

\begin{figure}[htbp]
    \centering
    \includegraphics[width=.7\textwidth]{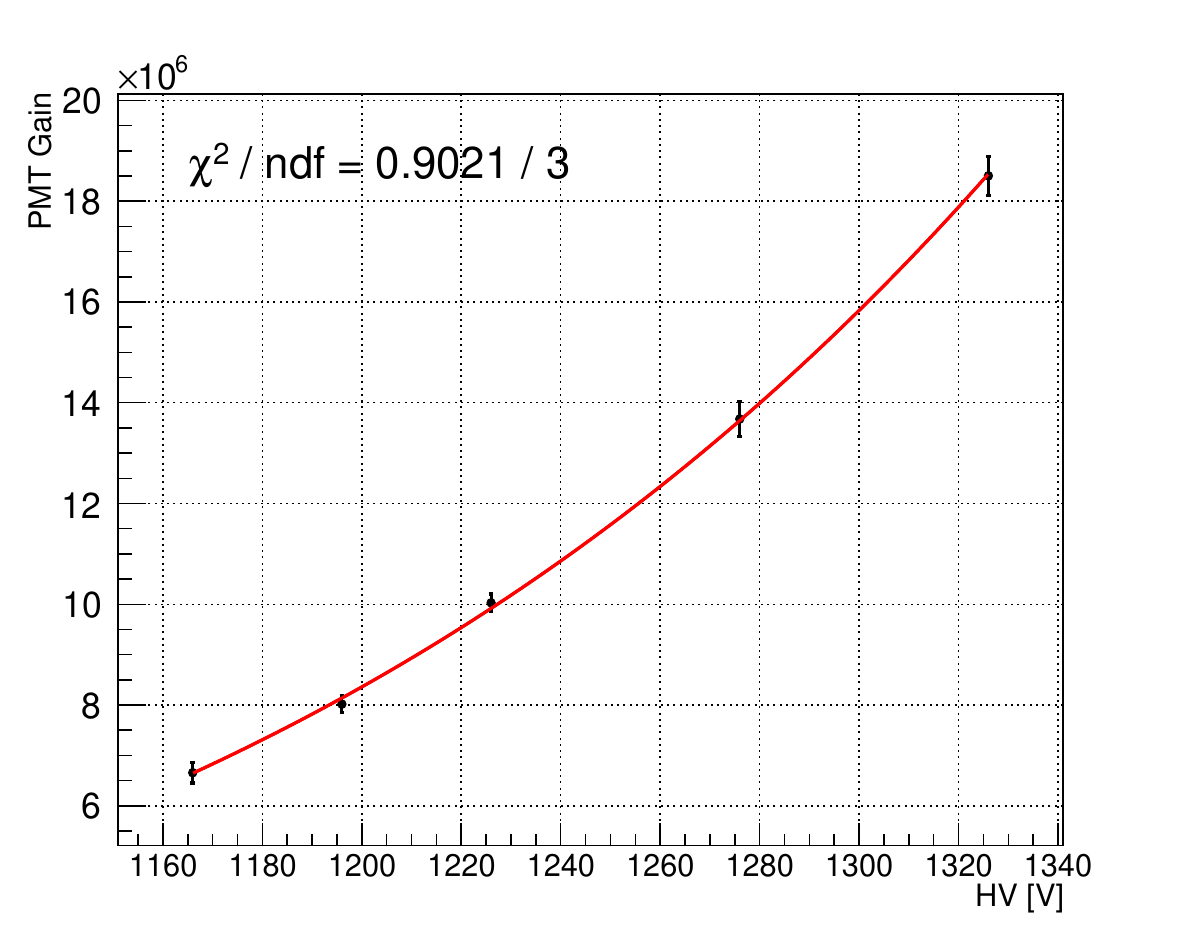}
    \caption{An example of collected charges in the PMT as a function of supplied HV. }
    \label{Fig:GC}
\end{figure}

The measured typical gain curve of a single PMT is shown in Figure~\ref{Fig:GC}. The single photoelectron peak positions were measured at varying high voltages (HVs), typically at four or five points, with 10,000 or 20,000 events collected for each HV value.
	
\begin{figure}[htb]
    \centering
    \includegraphics[width=.7\textwidth]{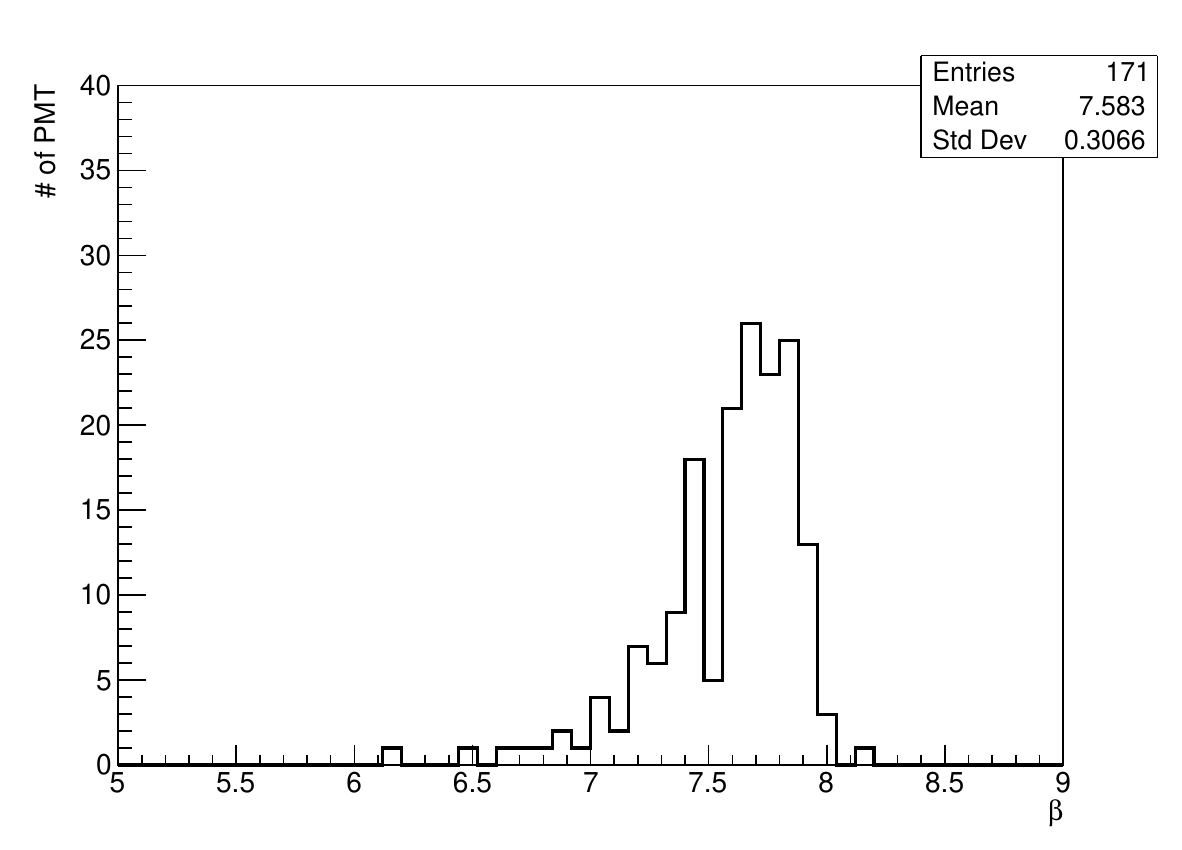}
    \caption{The beta-value distribution calculated based on Equation~\ref{eq:HVC} for the inner PMTs. }
    \label{Fig:beta}
\end{figure}

\begin{figure}[htb]
    \centering
    \includegraphics[width=.7\textwidth]{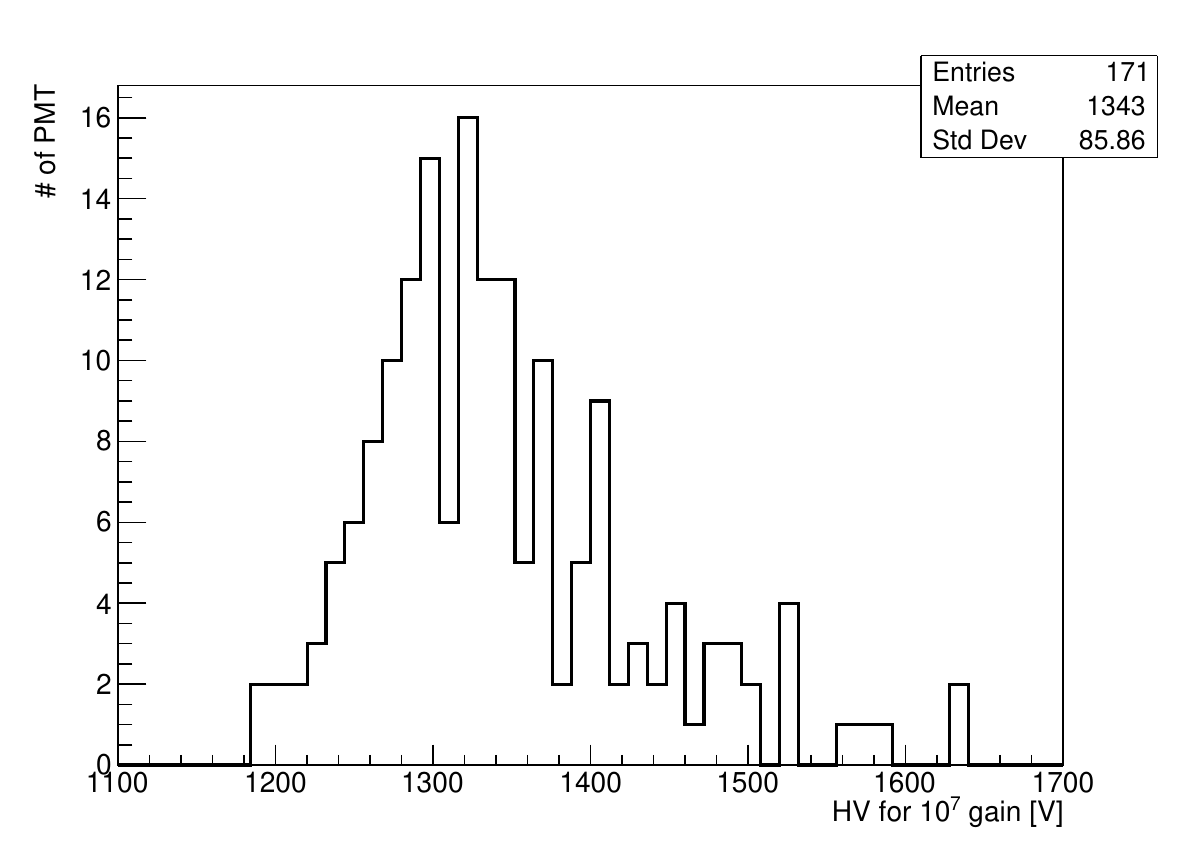}
    \caption{The high voltage (HV) values required to achieve a gain of 10$^{7}$ for the inner PMTs.}
    \label{Fig:10-to-7}
\end{figure}

To compare the curves to other measurements, the fit using Equation~\ref{eq:HVC} was 
performed.
\begin{equation}
   G = \alpha \times V^{\beta}.
   \label{eq:HVC}
\end{equation}
The fit was performed in each PMT. The $\beta$ values for the inner PMTs are shown in Figure~\ref{Fig:beta}, excluding one dead PMT. Additionally, Figure~\ref{Fig:10-to-7} shows the HV values required to achieve a gain of 10$^{7}$ for the inner PMTs.
Approximately 1500 V is needed to reach the 10$^{7}$ gain level.
To get a good energy resolution for physics events, a 10\% relative gain adjustment is required for the inner PMTs, and it will be achieved using measured HV curves and the LED calibration.

\begin{figure}[htb]
    \centering
    \includegraphics[width=.7\textwidth]{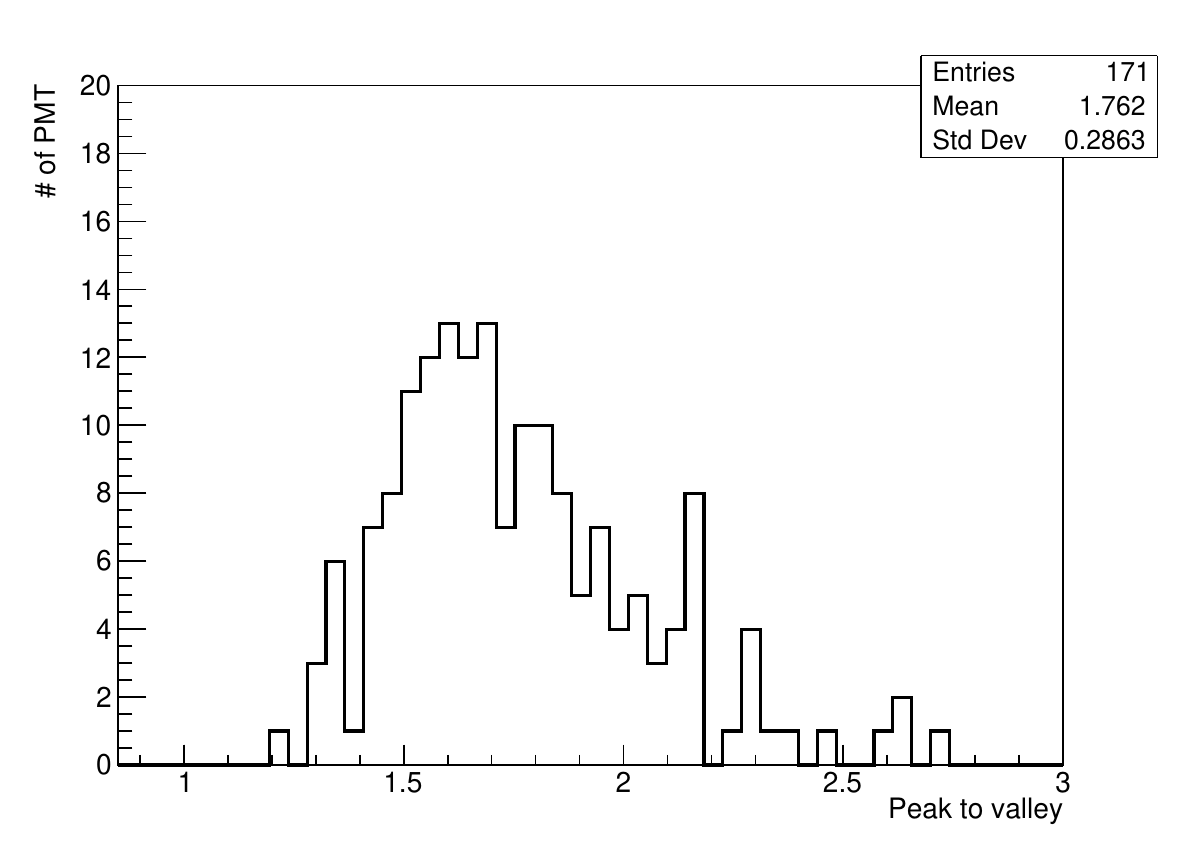}
    \caption{The Peak-to-Valley ratio of the inner PMTs.}
    \label{Fig:PV}
\end{figure}

The Peak-to-Valley (PV) ratio is defined as the ratio between the single photoelectron Gaussian peak and the lowest bin value between the pedestal and the single photoelectron peak. A higher PV ratio typically indicates better separation of the real signal from noise, including the baseline level. Figure~\ref{Fig:PV} shows the PV ratios for the inner~PMTs, 
with a mean value of approximately 1.762.
The Peak-to-Valley ratio values are similar to those from the production company, which distinguish the single photoelectron signal from the pedestal well.

\subsection{Relative T$_{0}$ measurement}

\begin{figure}[htbp]
		\centering
		\includegraphics[width=.7\textwidth]{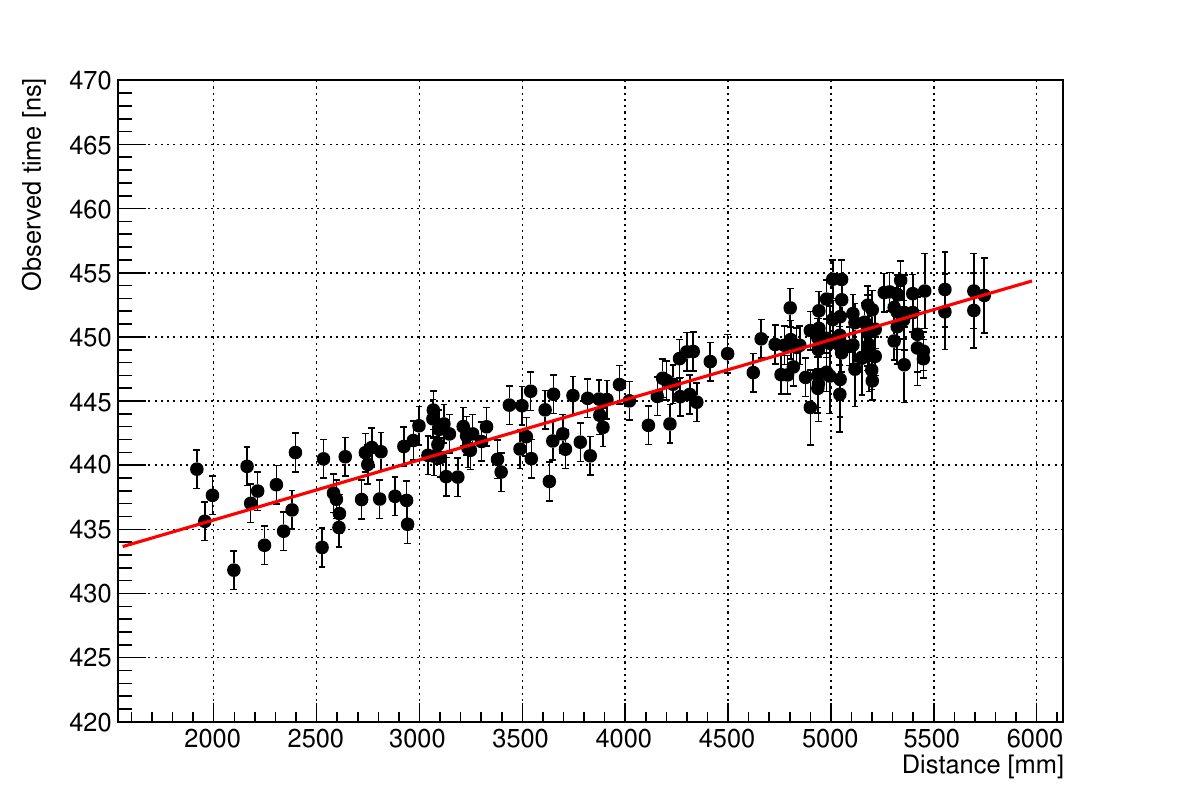}
		\caption{The observed pulse timings as a function of distance between PMT and LED.}
		\label{Fig:TOF}
	\end{figure}

Relative time (T$_{0}$) refers to the time interval from the trigger of the LED controller box to the PMT signal and is given by:  
\begin{equation}
    \mathrm{T_{0}} =
    \mathrm{T_{cable}}
    +\mathrm{T_{TOF}} 
    +\mathrm{T_{electronics}},
\end{equation}
where $\mathrm{T_{cable}}$ represents the delay due to the PMT cable length, $\mathrm{T_{TOF}}$ corresponds to the flight time of light from the LED to the PMT, and $\mathrm{T_{electronics}}$ accounts for the delay introduced by the associated electronics. The observed pulse timing is defined as the moment when the pulse exceeds the threshold.
%The pulse timing in each PMT is influenced by the PMT position, cable length, and associated electronics. 
Figure~\ref{Fig:TOF} shows the observed pulse timing as a function of the flight length. Note that the absolute values of observed time have an offset, which is an arbitrary value from the data acquisition system. The fitted result of a linear function corresponds to ($2.1\pm0.1)\times10^{8}$~m/s, consistent with the light velocity in the liquid scintillator, $2.0\times10^{8}$~m/s. The measurement achieves a precision within 2.1~ns, meeting the 10~ns requirement for event reconstruction in JSNS$^{2}$-II.
The measured relative time differences will be considered in the event reconstruction for the physics run of JSNS$^{2}$-II.

\section{Summary}

This paper describes the in-situ PMT calibration using an LED system. In particular, the UV LED produces 
pseudo-isotropic light in the liquid scintillator via wavelength shifters, which is useful for PMT gain calibration.
Key parameters such as gains, HV curves, PV ratios, and relative T$_{0}$ values were measured and utilized for real operation.
To get a good energy resolution for physics events, a 10\% relative gain adjustment is required for all functioning inner~PMTs, and it will be achieved using measured HV curves and the LED calibration. The peak-to-valley ratio values are similar to those from the production company, which distinguish the single photoelectron signal from the pedestal well.
Furthermore, the PMT signal timing accuracy was determined to be 2.1~ns, which meets the 10~ns requirement for event reconstruction.

\acknowledgments
%We thank the J-PARC staff for their support. We acknowledge the support of the Ministry of Education, Culture, Sports, Science, and Technology (MEXT) and the JSPS grants-in-aid: 16H06344, 16H03967 and 20H05624, Japan. 
%This work is also supported by the National Research Foundation of Korea (NRF): 2016R1A5A1004684, 2017K1A3A7A09015973, 2017K1A3A7A09016426,
%2019R1A2C3004955, 2016R1D1A3B02010606, 2017R1A2B4011200, 2018R1D1A1B07050425,
%2020K1A3A7A09080133, 2020K1A3A7A09080114, 2020R1I1A3066835, 2021R1A2C1013661 and 2022R1A5A1030700. Our work has also been supported by a fund from the BK21 of the NRF. The University of Michigan gratefully acknowledges the support of the Heising-Simons Foundation. This work conducted at Brookhaven
%National Laboratory was supported by the U.S. Department of Energy under Contract DE-AC02- 98CH10886. The work of the University of Sussex is supported by the Royal Society grant no. IESnR3n170385. We also thank the Daya Bay Collaboration for providing the Gd-LS, the RENO collaboration for providing the LS and PMTs, CIEMAT for providing the splitters, Drexel University for providing the FEE circuits and Tokyo Inst. Tech for providing FADC boards.

 We deeply thank the J-PARC for their continuous support, 
especially for the MLF and the accelerator groups to provide 
an excellent environment for this experiment.
We acknowledge the support of the Ministry of Education, Culture, Sports, Science, and Technology (MEXT) and the JSPS grants-in-aid: 16H06344, 16H03967, 23K13133, 
24K17074 and 20H05624, Japan. This work is also supported by the National Research Foundation of Korea (NRF): 2016R1A5A1004684, 17K1A3A7A09015973, 017K1A3A7A09016426, 2019R1A2C3004955, 2016R1D1A3B02010606, 017R1A2B4011200, 2018R1D1A1B07050425, 2020K1A3A7A09080133, 020K1A3A7A09080114, 2020R1I1A3066835, 2021R1A2C1013661, NRF-2021R1C1C2003615, 2021R1A6A1A03043957, 2022R1A5A1030700, RS-2023-00212787 and RS-2024-00416839. Our work has also been supported by a fund from the BK21 of the NRF. The University of Michigan gratefully acknowledges the support of the Heising-Simons Foundation. This work conducted at Brookhaven National Laboratory was supported by the U.S. Department of Energy under Contract DE-AC02-98CH10886. The work of the University of Sussex is supported by the Royal Society grant no. IESnR3n170385. We also thank the Daya Bay Collaboration for providing the Gd-LS, the RENO collaboration for providing the LS and PMTs, CIEMAT for providing the splitters, Drexel University for providing the FEE circuits and Tokyo Inst. Tech for providing FADC boards.

	% Bibliography
	
	%% [A] Recommended: using JHEP.bst file
	%% \bibliographystyle{JHEP}
	%% \bibliography{biblio.bib}
	
	%% or
	%% [B] Manual formatting (see below)
	%% (i) We suggest to always provide author, title and journal data or doi:
	%% in short all the informations that clearly identify a document.
	%% (ii) please avoid comments such as "For a review'', "For some examples",
	%% "and references therein" or move them in the text. In general, please leave only references in the bibliography and move all
	%% accessory text in footnotes.
	%% (iii) Also, please have only one work for each \bibitem.

\end{document}